\documentstyle[11pt]{article}

\newcommand{\leuven}[2]{#2}
\newcommand{\hep}[1]{#1}

\newcommand{\eqn}[1]{(\ref{#1})}
\def\W52{$W_{5/2}$}
\def\ie{{\sl i.e.\ }}
\def\eg{{\sl e.g.\ }}
\def\rhs{{\sl rhs}}

\def\WA{$W$--al\-ge\-bra}
\def\WS{$W$--string}
\def\be{\begin{equation}}
\def\ee{\end{equation}}
\def\bea{\begin{eqnarray}}
\def\beastar{\begin{eqnarray*}}
\def\eea{\end{eqnarray}}
\def\eeastar{\end{eqnarray*}}
\def\nonu{\nonumber \\}

\newcommand{\del}{\partial}
\newcommand{\bd}{{\bar \partial }}

\def\SBV{S_{BV}}

\def\ft#1#2{{\textstyle{{#1}\over{#2}}}}

\leuven{
\renewcommand{\section}[1]{\addtocounter{section}{1}
\vspace{5mm} \par \noindent
  {\bf \thesection . #1}\setcounter{subsection}{0}
  \par
   \vspace{2mm} } 

\renewcommand{\subsection}[1]{\addtocounter{subsection}{1}
\vspace{2.5mm}\par\noindent {\em \thesubsection . #1}\par
 \vspace{0.5mm} }
\renewcommand{\thebibliography}[1]{ {\vspace{5mm}\par \noindent{\bf
References}\par \vspace{2mm}}
\list
 {\arabic{enumi}.}{\settowidth\labelwidth{[#1]}\leftmargin\labelwidth
 \advance\leftmargin\labelsep\addtolength{\topsep}{-4em}
 \usecounter{enumi}}
 \def\newblock{\hskip .11em plus .33em minus .07em}
 \sloppy\clubpenalty4000\widowpenalty4000
 \sfcode`\.=1000\relax \setlength{\itemsep}{-0.4em} }
}{
}

\begin{document}

\begin{flushright} Preprint KUL-TF-95/22 \\
                   Preprint Imperial/TP/94-95/58
\hep{                \\   hepth@xxx/9511200}
\end{flushright}

\vspace{4mm}

\begin{center}
{\bf GAUGING CONFORMAL ALGEBRAS WITH RELATIONS BETWEEN THE GENERATORS
\leuven{
  \footnote{This talk (presented by K.~Thielemans) was also given at
  the Sixth Seminar on Quantum Gravity, Moscow, 12-17 June '95.}
}{}
\hep{
  \footnote{This paper is based on proceedings for
  the Sixth Seminar on Quantum Gravity, Moscow, 12-17 June '95
  and the Conference on Gauge theories, Applied Supersymmetry and Quantum
  Gravity, Leuven, 10-14 July '95.}
}
}
\vspace{1.4cm}

K.~THIELEMANS \footnote{E-mail: k.thielemans@ic.ac.uk}\\
{\em Theoretical Physics Group, Imperial College\\
London SW7 2BZ, UK} \\
\vskip .5cm
S.~VANDOREN\footnote{E-mail: stefan.vandoren@fys.kuleuven.ac.be} \\
{\em Instituut voor Theoretische Fysica, K.U.Leuven
        \\Celestijnenlaan 200D, B-3001 Leuven, Belgium}\\[0.3cm]
\end{center}

\centerline{ABSTRACT}
\vspace{- 4 mm}
\begin{quote}\small
We investigate the gauging of conformal algebras with relations between
the generators.
We treat the $W_{5/2}$--algebra as a specific example. We show
that the gauge-algebra is in general reducible with an infinite number of
stages. We show how to construct the BV-extended action, and hence the
classical BRST charge. An important conclusion is that this can always
be done in terms of the generators of the $W$--algebra only,
that is, independent of the realisation.\\
The present treatment is still purely classical, but already enables
us to learn more about reducible gauge algebras and the BV-formalism.
\end{quote}
\vspace{2mm}
\normalsize

\section{Introduction}
It cannot be stressed enough that gauge symmetries play an extremely
important role in our understanding of particle physics. Therefore it is
very important to study the quantisation of models possessing a number of
gauge invariances. By now, we know there is a large variety of these models,
from electromagnetism, Yang--Mills and gravity theories to supergravity,
$W$-gravities, superparticles and superstrings. All these models can, apart
from their field content, be characterised by their {\it algebra of gauge
transformations}. In Yang--Mills, this algebra is a Lie-algebra with
structure constants satisfying Jacobi identities. In supergravity theories
one has to extend this to more general gauge algebras, where one can
have structure functions and where the
algebra only closes modulo (graded) antisymmetric combinations of
the field equations
\cite{Kallosh,deWitvH}. For these so
called open gauge algebras, one cannot simply apply the same BRST
quantisation method as for Yang--Mills theory, and an appropriate extension
of the BRST formalism was given in \cite{deWitvH}.
In the case where the gauge symmetries are not independent, the level of
reducibility is another important characterisation.
As an example of a first
level reducible theory, one can think about the antisymmetric tensor
\cite{Siegel} where one has to introduce, on top of the ordinary ghosts, a
ghost for ghosts. Other more complicated examples are the superparticle and
the Green--Schwarz superstring, which are infinite stage reducible theories.
In these cases, one has to work with an infinite tower of ghost for ghosts,
see \eg \cite{superparticlestring}.

In this paper, we will even go one step further. We will start with an
action $S_0[\phi^i]$ with a number of global symmetries with generators
$T_a$, which are then gauged by introducing gauge fields $\mu^a$~:
\begin{equation}
S=S_0[\phi] +\int\mu^a T_a(\phi )\ ,
\end{equation}
where we call $\phi ^i$ the matter fields.  We will concentrate on a
specific example of a two dimensional conformal field theory based on the
(nonlinear) \W52-- algebra with the Virasoro spin 2 current $T_1=T$ and a
spin 5/2 fermionic current $T_2=G$. The new thing in this model is that the
gauge algebra {\it does not close} on the two gauge symmetries, even when
using antisymmetric combinations of field equations~! Instead, it generates
2 new unexpected symmetries
(on shell zero) that act only on the gauge fields. It turns out that we
have to include these 2 symmetries to find a gauge algebra that closes up
to trivial symmetries (antisymmetric combinations of field equations).
However, as we will show, adding these new symmetries to the original ones
will make the complete set of symmetry generators dependent, so that we are
dealing with a reducible theory. After introducing the necessary zero modes
and their corresponding ghosts for ghosts, we will even see that the
complete set of zero modes is reducible itself. This is a never ending
story~: the reducibility has an infinite number of stages, and there is an
infinite tower of ghosts.

All this can be better understood in terms of ``nonfreely generated''
conformal algebras. These are algebras where the Jacobi identities are
only satisfied if certain combinations of the generators are considered
to be zero (``null fields''). The simplest example, which we also
consider in this paper, is the \W52--algebra, discovered in the quantum
case in \cite{zamoW3}. It is then clear that, because there are
relations between the currents $T_a$, there will be extra symmetries in
the theory \cite{BRT}. They are precisely those needed to close the
gauge algebra. Together with the original ones, they will form a
reducible system \cite{BRT,superparticlestring}. It is the gauge theory
of this conformal algebra that we want to treat here.

To handle such a complicated system, we resort to the antifield formalism of
Batalin--Vilkovisky (BV) \cite{BV}. We sketch how to deal with
further zero modes that vanish on shell, a point that is not well discussed
in the literature so far. To do this, we make use of the acyclicity of
the Koszul--Tate differential, the basic ingredient of the (BV) formalism.
Details are given in \cite{more}.

The main motivation for this work lies in the further study of gauge
algebras. The gauging of nonfreely generated \WA s is however
interesting in its own right, as this could provide a new class of \WS\
theories. Indeed, up to now, all \WS\ theories are constructed by gauging a
\WA\ where all generators are linearly independent. Furthermore,
a particular class of nonfreely generated quantum \WA s have been studied
lately. They provide ``unifying'' \WA s for the more familiar algebras in the
Drinfeld--Sokolov series \cite{unifyWA}. The study of \WS s based on the
unifying algebras will however be complicated by the fact that the classical
versions of these \WA s have an infinite number of generators. Clearly, we
first have to understand the case of nonfreely, but finitely generated
\WA s.

So, in the next section we study the \W52 current algebra, and
discuss how the relations between the currents follow from the Jacobi
identities. Then, in section 3 we show how the extra symmetries are
generated starting from the (open) gauge algebra based on the
gauged (super)conformal symmetries. In section 4
we show that the model is infinitely reducible. In a last section,
we discuss briefly the gauge fixing procedure in the BV formalism
and determine the structure of the BRST charge. We end with some
conclusions.

\section{The current algebra}
The \W52--algebra was one of the first
\WA s constructed, see \cite{zamoW3} where it is presented in the
quantum case with Operator Product Expansions. We need it here as a
classical \WA, \ie using single contractions. The algebra consists of two
currents~: $T$ the Virasoro generator and a primary dimension $\ft52$
current $G$. They satisfy the brackets~:
\bea
\{T(z),T(w)\}&=& -2T(w) \del\delta(z-w)+\del T(w) \delta(z-w)\nonu
\{T(z),G(w)\} &=& -\ft 5 2 G(w) \del\delta(z-w)+\del G(w) \delta(z-w)\nonu
\{G(z),G(w)\} &=& T^2(w) \delta(z-w)\ .\label{PBalgebra}
\eea
The last bracket leads us to call $G$ a (generalised) supersymmetry
generator.
In the quantum case, the Jacobi identities are only satisfied for a
specific value of the central charge $c=-\ft{13}{14}$ and even then only
modulo a ``null field''.
In this context, we call ``null fields'' all the combinations of $T$ and
$G$ which should be put to zero such that the Jacobi identities are
satisfied. Similarly, we find in the classical case that
the algebra does not admit a central extension and there is a
classical null field~:
\be
N_1\equiv 4T\, \del G -5 \del T\, G \ . \label{nullfield}
\ee
We can check by repeatedly computing brackets with
$N_1$ that the null fields are generated by $N_1$ and
\be
N_2 \equiv 2 T^3 - 15 \del G\, G\ . \label{nullfield2}
\ee
More precisely, all other null fields are of the form~:
\be
f_n(T,G)\ \del^n N_1 + g_m(T,G)\ \del^m N_2
\ee
where $f_n,g_m$ are differential polynomials in $T$ and $G$.

A realisation for the algebra \eqn{PBalgebra} was found in \cite{TAMhs}~:
\bea
T&=& -\ft12 \psi\, \del\bar\psi +\ft12 \del\psi\, \bar\psi \ ,\nonumber\\
G&=& \ft12\left(\psi+ \bar\psi\right)\, T \ ,
\label{W52realisation}
\eea
where $\psi$ is a complex fermion satisfying the Dirac brackets
$\{\psi(z),\bar\psi(w)\}=\delta(z-w)$.  One can easily verify for this
realisation that the null fields $N_i$ vanish.

In fact, for {\sl any} realisation in terms of fields of an underlying
Poisson (or Dirac) algebra (\eg free fields), the null fields will vanish
identically. Indeed, they appear in the \rhs\ of a Jacobi identity,
which is of course satisfied for a Poisson algebra. This means
that in any realisation, the generators $T,G$ are not independent. They
satisfy (at least) the relations $N_i=0$. In the following section, we will
see that these relations have important consequences for the gauge algebra.

\section{The gauge algebra}
In order to construct a gauge theory based on this algebra, one must
be able to work in a certain realisation, \ie one must specify an action
$S_0$ for matter fields $\phi^i$
\footnote{For the complex
fermion in \eqn{W52realisation}, one has $S_0=\psi{\bar\partial}\bar\psi$.}.
Using this action $S_0$, one can define light--cone Poisson (or Dirac)
brackets between the
fields and their momenta. With respect to these brackets, we assume that
we can find conserved currents $T(\phi )$, $G(\phi )$
satisfying the algebra \eqn{PBalgebra}. The
transformations of the fields are obtained by taking
brackets with the generators~:
\begin{equation}
\delta_{\epsilon^a} \phi = \int \epsilon^a \{ T_a, \phi \}
\label{phitrans}
\end{equation}
where the index $a$ runs over the number of generators, and there is no
summation on the \rhs . We will not make a choice for the realisation and
use only the information contained in the algebra of the generators to
construct a gauge theory.

The above assumption implies that the action $S_0$ transforms
under the conformal symmetry and supersymmetry with parameters
$\epsilon $ and $\alpha $ respectively, as
\begin{equation}
\delta _\epsilon S_0=
  -\int \bd\epsilon T \qquad  \delta_\alpha S_0=-\int\bd\alpha G\ ,
\label{Noether}
\end{equation}
where the transformations of the Noether currents $T,G$ follow from
eq.~\eqn{phitrans}
\leuven{.

}{~:
\begin{eqnarray}
\delta _\epsilon T=\epsilon \partial T+2\partial \epsilon T&\qquad&
\delta _\alpha T=\frac{3}{2}\alpha \partial G+\frac{5}{2}\partial \alpha
G \nonumber\\
\delta _\epsilon G=\epsilon \partial G+\frac{5}{2}\partial \epsilon
G&\qquad& \delta _\alpha G=\alpha T^2
\label{TGtransform}
\end{eqnarray}
}
The commutators between two symmetries can be computed using the Jacobi
identities:
\begin{eqnarray*}
[\delta_{\epsilon^a}, \delta_{\tilde\epsilon^b}] \phi &=&
  \int \epsilon^a \int \tilde\epsilon^b
  \left((-1)^{ab}\{ T_a, \{ T_b, \phi \}\} - \{T_b, \{T_a,\phi \}\}\right)\\
&=&
  -\int \epsilon^a \int \tilde\epsilon^b
  \{ \{ T_b,  T_a\}, \phi \} \ .
\end{eqnarray*}
We find~:
\begin{eqnarray}
{[}\delta_{\epsilon _1},\delta _{\epsilon _2}] &=&
   \delta _{\tilde\epsilon=\epsilon _2\partial \epsilon _1-
        \epsilon _1\partial \epsilon _2}\nonumber\\
{[} \delta _\epsilon,\delta _\alpha ] &=&
   \delta _{\tilde \alpha=
   -\epsilon \partial \alpha +3/2\alpha \partial \epsilon}
    \nonumber\\
{[} \delta _{\alpha _1},\delta _{\alpha _2}] &=&
   \delta _{\tilde\epsilon = 2\alpha _2\alpha _1T}\,.
\label{chiralcommutators}
\end{eqnarray}
Now, we can gauge these symmetries by introducing gauge fields
$\mu $ (bosonic) and $\nu $ (fermionic) for the
conformal and susy symmetries. The action is then
\begin{equation}
S=S_0 +\int\mu T+\int\nu G\ .\label{fullS}
\end{equation}
The transformation rules
for the gauge fields such that the action is invariant, are
\begin{eqnarray}
\delta _\epsilon \mu =\nabla ^{-1}\epsilon&\qquad &
\delta _\alpha \mu =\alpha \nu T\nonumber\\
\delta _\epsilon \nu =\epsilon \partial \nu -\frac{3}{2}\nu \partial
\epsilon &\qquad&
\delta _\alpha \nu =\nabla ^{-(\frac{3}{2})}\alpha \ ,\label{gtr}
\end{eqnarray}
with the notation $\nabla ^j=\bd-\mu \partial -j\partial \mu $. These
rules enable us to study the gauge algebra. Computing the commutators of
the gauge symmetries on the gauge fields, we see that they close only
after using equations of motion.

In the usual case for open algebras
\cite {deWitvH} one has the following structure~:
\begin{equation}
[\delta _{\epsilon ^a},\delta _{\epsilon ^b}]\phi
^A=R^A_cT^{c}_{ab}\epsilon ^b\epsilon ^a-y_BE^{BA}_{ba}\epsilon
^a\epsilon ^b\ ,\label{openalg}
\end{equation}
where $\phi ^A$ now stands for all the classical fields (matter fields
$\phi^i$ and gauge fields $\mu^a$),
and the symmetries are written in the form $\delta_\epsilon \phi
^A=R^A_a\epsilon ^a$. The structure functions $T^a_{bc}$ are
graded antisymmetric in $(bc)$. The first term in the \rhs\ can be
rewritten as $\delta _{{\tilde \epsilon }}\phi ^A$ for ${\tilde \epsilon
}^c=T^c_{ab}\epsilon ^a\epsilon ^b$, so this is again a symmetry of the
action. The second term is proportional to field equations $y_A$. As it
arises in the commutator of two symmetries, it leaves the action invariant
too. This is trivially the case when the matrix $E^{AB}=E^{AB}_{ab}\epsilon
^a\epsilon ^b$ is graded antisymmetric in $(AB)$, because then it generates
trivial field equation symmetries of the form $\delta \phi ^A=E^{AB}y_B$,
and one does not need to take these into account for quantising the theory.
All previously known examples of gauged algebras are of the type
\eqn{openalg}, with a graded antisymmetric $E^{AB}$-matrix
and $R^A_a$ on shell nonzero.

In the case of \W52 however, the commutator of two supersymmetries gives
us something unexpected. Computing the commutator
\eqn{chiralcommutators} on the gauge fields, we find~:
\begin{eqnarray}
\Bigl(\left[ \delta _{\alpha _1},\delta _{\alpha _2}\right] -
\delta _{\tilde\epsilon = 2\alpha _2\alpha _1T}\Bigr) \mu
&=& -[\nabla^{-3}(\alpha _2\alpha _1)T+2\alpha _2\alpha
_1\nabla^{2}T]\nonumber\\
&&-[\frac{5}{2}\partial (\alpha _2\alpha _1)\nu G +3\alpha _2\alpha _1\nu
\partial G]\nonumber\\
\Bigl( \left[ \delta _{\alpha _1},\delta _{\alpha _2}\right] -
      \delta _{\tilde\epsilon = 2\alpha _2\alpha _1T} \Bigr) \nu
&=&-[3
\alpha _2\alpha _1\partial (T\nu )+\frac{1}{2}\partial (\alpha _2\alpha
_1)\nu T]\nonumber\\
&&+9\alpha _2\alpha _1\nu \partial T+\partial (\alpha _2\alpha _1\nu )T\ ,
\end{eqnarray}
upon using (\leuven{}{\ref{TGtransform},}\ref{gtr}).
All the terms in square brackets can be absorbed by trivial field
equation symmetries
using the field equations $y_\mu=T$ and $y_\nu=-G$.
\footnote{The field equations are defined as
the right derivative of the action w.r.t. the field.}.
However, the two terms on the last line of the \rhs\ for $\nu$ remain.
So, they come from a nontrivial symmetry, which is zero on shell~:
\begin{eqnarray}
\delta_{n} \phi^i\ =\ 0\qquad
\delta_n\mu\ =\ 0\qquad
\delta_{n} \nu\ =\ 9n\partial T+4\partial n T\ .\label{eps1gtr}
\end{eqnarray}
Note that it acts only on the gauge fields, and hence leaves $S_0$
invariant. The variation of the action \eqn{fullS} under these
transformation rules is proportional to the relation $N_1$
\eqn{nullfield}. Of course, in hindsight it is obvious there is a
corresponding symmetry associated with such a relation. However, if one
tries to quantise the action without knowing the algebra of the previous
section, one is surprised that the gauge transformations \eqn{gtr} do
not form a closed algebra, even after using trivial equation of motion
symmetries.

Completely analogously, one can find a second new symmetry. This symmetry
will appear in the commutator $[\delta _\alpha ,\delta _n]$, again
acting on the
gauge fields.
The second
symmetry, with bosonic parameter $m$ can be written as
\begin{eqnarray}
\delta _m\phi^i\ =\ 0\qquad
\delta _m\mu \ =\ 2m T^2\qquad
\delta _m\nu\ =\ -15m\partial G\ , \label{eps2gtr}
\end{eqnarray}
which indeed leaves the action \eqn{fullS} invariant when using $N_2=0$
\eqn{nullfield2}.

The two new symmetries are proportional to field equations. The way
they are written down is not unique. For instance, one could change
\eqn{eps1gtr} to $\delta _n \mu=-4n \partial G; \delta
_n \nu =5n \partial T$. This choice is however
equivalent, since it corresponds to \eqn{eps1gtr} by adding a trivial
field equation symmetry, and this does not change the theory and its
quantisation.

\section{Reducibility}
Having found all the gauge symmetries, we should see if they are all
independent, \ie is the gauge algebra irreducible ? We have to
investigate if we can find zero modes $Z$ of the matrix $R$ of gauge
transformations~:
\bea
R^A_a Z^a_{a_1} \epsilon^{a_1} = y_B f^{BA}\ , \label{zero modes}
 \eea
where the index $a$ runs over all symmetries ($1\ldots 4$).
Remark that the $Z^a_{a_1}$ are only zero modes on the stationary surface,
and the $f^{BA}$ are graded antisymmetric, see e.g. \cite{BVles}.
We expect that these zero modes
will be related to the relations $N_i=0$ \cite{BRT}. Let us first look at the
transformations of the matter fields $\phi^i$. Consider the
transformations generated by taking Poisson brackets with the $N_i$.
Because the relations contain only the generators $T,G$, we can use the
Leibniz rule and some partial integrations to rewrite the transformation
as a combination of those generated by $T,G$. For example~:
\bea
\int \zeta^1 \{N_1, \phi^i\} &=&
\left( \delta_{\epsilon=9 \zeta^1\del G  + 5 \del \zeta^1 G}-
       \delta_{\alpha=9 \zeta^1\del T  + 4 \del \zeta^1 T}
         \right) \phi^i \ .\label{symrelation1}
\eea
However, because $N_1=0$, the previous equation gives us a relation
between the transformations of the matter fields (valid for every
realisation). Similarly, via $N_2$ we can find another relation between
the gauge transformations acting on the matter fields.

These two relations satisfy eq.~\eqn{zero modes} for the $A$--index running
over the matter fields (with $f^{Bi}=0$). However, for
the gauge fields $\mu,\nu$, we find that we need the
extra symmetries eqs.~(\ref{eps1gtr},\ref{eps2gtr}) to make $f$
graded antisymmetric. Of course, we can
include these terms in eq.~\eqn{symrelation1}
as the extra symmetries do not act on the matter fields.
Summarising, we find two zero modes
(one for every relation $N_i$)
giving the following entries in the matrix $Z^a_{a_1}$~:
\be
\begin{array}{c|c}
-9 \del G - 5 G \del  & -6 T^2\\
-9 \del T - 4 T \del  & 30 \del G + 15 G \del\\
\nabla^{-{9\over 2}} & -{1\over 2} \nu T\\
{3\over 2} \del \nu -{1\over 2}\nu \del & \nabla^{-5}
\end{array} \label{Z1a}
\ee
where the rows correspond to the conformal symmetry, the
supersymmetry and the two extra symmetries (\ref{eps1gtr},\ref{eps2gtr}),
and each column corresponds to a zero mode.

Surprisingly, this is not the end of the story. When trying to construct
a BRST charge in the BFV formalism, or an extended action in BV, no
solution can be found with the above symmetries and zero modes.
Indeed, many other zero modes exist. They all have zero entries in the first
two rows of $Z$, that is, they are relations between the extra symmetries.
Furthermore, the remaining two entries are differential polynomials in
$T,G$, which means that they are zero on shell. We give as an example
the zero mode $Z = (0, 0, T, 0)^t$, for which there indeed exists a
graded antisymmetric $f^{AB}$ such that eq.~\eqn{zero modes} is satisfied.

However, most of these zero modes do not solve the problems mentioned
above.  In fact, we have to look more closely at the existence proof of the
relevant object (BRST-charge or extended action). Both proofs involve the
computation of the cohomology of the so--called Koszul--Tate differential
$\delta_{KT}$ \cite{KT}. We do not wish to go into details here (see
\cite{more}), but give only the gist of the argument.

Eq.~\eqn{zero modes} corresponds to the existence of a KT--invariant (or
cocycle). However, only those invariants which are not exact
--- \ie not the $\delta_{KT}$ of something else ---
determine the non trivial zero modes.
In the BV
language, this can be stated as follows. If the BV master equation cannot
be satisfied at a certain antifield level, it is because a
KT--nontrivial cocycle $A$ exists. One then introduces a cochain
$a$ by hand, such that $\delta_{KT}a = A$, making $A$ exact. In our case,
the cochain $a$ would be
the antifield of a ghost for ghost.

All this means we have to compute the cohomology of $\delta_{KT}$ (at
antifield level 2). One can do this for cocycles organised by engineering
\footnote{The engineering dimension is
minus the dimension in meters. $\del$ and $\bd$ increase it by one.}
and conformal dimension.
The result of this calculation in our present case
is that $(0,0,T,0)^t$ corresponds to a trivial zero mode, \ie it
is $\delta _{KT}$ exact without introducing a new ghost for ghost.
In fact, we need the
two following zero modes for which we give
only the two bottom rows of $Z$ (the first two contain zeroes)~:
\be
\begin{array}{c|c}
0       & T^2\\
\del T  & 0
\end{array} \label{Z1b}
\ee
So, we find that the acyclicity of the KT--differential implies the
introduction of zero modes which vanish on shell. To our knowledge, this
is the first algebra where this has been observed.
See \cite{more} for
more details. It is not clear to us how the zero modes \eqn{Z1b}
relate to the Poisson algebra of $T,G$.

With the information in (\ref{Z1a},\ref{Z1b}), we can continue the
computation of the extended action one step further. At the next level, we
have to look for zero modes of the $Z$ matrix~:
\bea
Z^{a}_{a_1} Z^{a_1}_{a_2} \epsilon^{a_2} = y_B f^{Ba}\ .
\label{zerozero modes}
 \eea
Again, this is only a necessary condition for the elements of $
Z^{a_1}_{a_2}$, and we have to compute the cohomology of $\delta_{KT}$
(now at antifield level 3) to see what the nontrivial zero modes are.
This calculation was done in Mathematica \cite{BVmat}.
We get the following table for $Z^{a_1}_{a_2}$
dropping the first two lines which contain only zeroes~:
\be
\begin{array}{c|c|c|c|c|c}
\nabla^{-8}                  & 0                 & 0     & T^2 & 0   & 0\\
4 \del\nu -{3\over 4}\nu\del &\nabla^{-{17\over 2}}& \del T& 0   & T^2 & T G
\end{array} \label{Z2}
\ee
It is now clear that we will find zero modes for $Z^{a_1}_{a_2}$ and so on.
This means that a gauged \W52 system is reducible with an infinite number of
stages.

\section{Gauge fixing}
In this section, we show briefly how the gauge fixing can be performed and
how the resulting BRST charge looks. In general, we need to
introduce ghosts $c^a$ for every symmetry $R^A_a$, ghosts for ghosts
$c^{a_1}$ for every zero mode $Z^a_{a_1}$ and so on. We will split
the ghosts in two classes. Ghosts for which there appears a $\nabla$
in $Z^{a_{i-1}}_{a_i}$, we denote by $c^{\{i\}}$,
the remaining ones by $\tilde{c}^{\{i\}}$.
The number of $\tilde c^{\{i\}}$ is equal to the number of $c^{\{i+1\}}$.


The gauge fixing is most easily done in the BV formalism \cite{BV}, or its
Hamiltonian counterpart. For details, see \cite{BVles}. One introduces
antifields for every field (including the ghosts) and forms the
extended action~:
\bea
\SBV &=& S + \phi^*_A R^A_a c^a +
    c_{a_i}^* Z^{a_i}_{a_{i+1}} c^{a_{i+1}} + \ldots
\eea
where the ellipsis denotes terms at least quadratic in antifields or ghosts.
They are determined by the (classical) master equation.

The gauge choice we take consists of putting the gauge fields $\mu,\nu$
and all the ghosts $\tilde{c}^{\{i\}}$ to zero. This can be done using
the symmetries associated with columns with $\nabla$ in $R$ and the
$Z$'s. In BV, this can be accomplished by performing a canonical
transformation transforming the antifields $\mu^*,\nu^*$ into the
antighosts $b_{\{1\}}$ and vice versa. Similarly, we transform
$\tilde{c}_{\{i\}}^*$ into $b_{\{i+1\}}$. The gauge fixed action
$S_{gf}$ is then obtained by putting the new antifields equal to zero.
It is nearly a free field action (due to the presence of the
$\nabla$'s), but there are additional terms like $b^{2}\bd c^1 c^1$. An
additional canonical transformation (similar to the one used for $W_3$
in \cite{StefanW3}) gives us a free field action for the ghosts~:
\be
S_{gf}= S_0 + b_{\{i\}} \bd c^{\{i\}}\ .\label{eq:Sgf}
\ee

Moreover, the extended action is linear in the new
antifields\footnote{This follows from dimensional arguments. In a
conformal field theory, we can associate two dimensions with every
field: the conformal dimension $d$ and $\bar d=D-d $, where $D$ is the
engineering dimension. All terms in the extended action need to have
$d=1,\bar d=1$. In our gauge choice, all fields have $\bar d= 0$ and
antifields have $\bar d=1$.}. This means that the BRST transformations
in this gauge choice are nilpotent off shell. The corresponding BRST
charge is~:
\be
Q = \oint c T + \gamma G +
(5 b\del G - 4 \del b G) r^1 + (15 b T^2 - 2 \beta \del G) r^2 +
\ldots\ ,
\ee
where we called the ghost of the conformal symmetry $c$ (antighost $b$),
of the supersymmetry $\gamma$ (antighost $\beta$), and the ghosts for
ghosts $r_1,r_2$.

\section{Discussion}
 From the example of the \W52--algebra, we can draw some general
conclusions for systems with relations between the generators and where
minimal coupling is sufficient for gauging.

In general, two kinds of relations are possible. The ones we studied
in this paper arise from the algebra, in particular from Jacobi
identities. These relations have to be satisfied for any realisation
having that particular symmetry algebra. On the other hand,
accidental relations
(valid in a particular realisation) are also possible.
An example is given by the realisation \eqn{W52realisation}
of the \W52 algebra.
There we have additional relations like~:
\bea
\psi \bar\psi T=0&& G - {1\over 2} (\psi+\bar\psi) T=0\ .
\label{accidental}
\eea
These relations explicitly involves matter fields.

We find an extra symmetry for every relation between the generators
\cite{BRT,superparticlestring}. These symmetries act only on the gauge fields.
If the relations involve only the generators the symmetries will be zero
on shell. If the
relations arise because of Jacobi identities, the extra symmetries have to be
included in the algebra \eqn{openalg} to make the matrix $E^{AB}$
graded antisymmetric. If there are only accidental relations, the
gauge symmetries that correspond to the global symmetries
form a subalgebra.
However, the extra symmetries do show up in the cohomology of $\delta_{KT}$.

The presence of the relations (and the extra symmetries) makes the gauge
algebra reducible. There is a zero mode for every relation \cite{BRT}.
By studying the cohomology of the Koszul--Tate differential, we can find
other zero modes.  If the relations consist of generators only, these
extra zero modes vanish on shell.  The zero modes then turn out to be
dependent themselves.

It is proven in \cite{KT} that for a class of theories, called ``regular'',
no on shell vanishing symmetries or zeromodes can occur. If we avoid using
a realisation, the \W52 --theory is an example of a nonregular theory.
Nevertheless, we showed that the BV formalism can still be applied by studying
the cohomology  of the Koszul-Tate differential, see also \cite{more}.
The realisation
\eqn{W52realisation} is regular however. The apparent contradiction with the
theorem of \cite{KT} is resolved by noting that now there are more extra
symmetries (which do not vanish on shell) corresponding to the relations
\eqn{accidental}. The symmetries (\ref{eps1gtr},\ref{eps2gtr}) are then
$\delta_{KT}$ trivial, and do not have to be included.

In contrast to the superparticle and superstrings, gauge
fixing does not present any problems, at least for \WA s.

It is surprising that we can start from the symmetries arising from only
the ``necessary'' relations, construct an extended action (or BRST
charge), and perform a valid gauge fixing.
For example, in the case of the realisation \eqn{W52realisation},
we did not include
the accidental symmetries \eqn{accidental}. Still,
the resulting gauge--fixed action
\eqn{eq:Sgf} does not have any remaining gauge
symmetries. This is related to the reducibility of the system.
It  would be interesting to investigate wether or not the theories
constructed using only necessary symmetries
are related to those where
all symmetries are gauge fixed.

An important question that remains is of course what happens
when quantising these systems. Are the extra symmetries anomalous ? We leave
this for further study.

\vspace{.3cm}
\noindent
{\bf Acknowledgements}\\
It is a pleasure to thank Jos\'e Figueroa-O'Farrill, Antoine Van Proeyen,
Ruud Siebelink, Jordi Par\'\i s, Kelly Stelle and Roya Mohayaee for helpful
discussions, and Lucy Wenham for numerous improvements of the manuscript.
\leuven{}{\\
This work was partly carried out in the framework of the project ``Gauge
theories, applied supersymmetry and quantum gravity'', contract
SC1-CT92-0789 of the European Economic Community.
}


\end{document}